# Machine learning complete intersection Calabi-Yau 3-folds

Kaniba Mady Keita[*]

*Centre de Calcul de Modélisation et de Simulation: CCMS; Department of Physics,*
*Faculty of Sciences and Techniques, University of Sciences, Techniques and Technologies of Bamako,*
*FST-USTTB, BP: E3206, Mali*

Gaussian process regression, kernel support vector regression, the random forest, extreme gradient boosting, and the generalized linear model algorithms are applied to data of complete intersection Calabi-Yau threefolds. It is shown that Gaussian process regression is the most suitable for learning the Hodge number $h^{2,1}$ in terms of $h^{1,1}$. The performance of this regression algorithm is such that the Pearson correlation coefficient for the validation set is $R^2 = 0.9999999995$ with a root mean square error $RMSE = 0.0002895011$. As for the train set, these two parameters are as follows: $R^2 = 0.9999999994$ and $RMSE = 0.0002854348$. The training error and the cross-validation error of this regression are $1 \times 10^{-9}$ and $1.28 \times 10^{-7}$, respectively. Learning the Hodge number $h^{1,1}$ in terms of $h^{2,1}$ yields $R^2 = 1.000000$ and $RMSE = 7.395731 \times 10^{-5}$ for the validation set of the Gaussian process regression.

## I. INTRODUCTION

String theory is currently known as a powerful candidate of quantum gravity. However, consistency of string theory requires 26-dimensional Minkowski spacetime for bosonic strings, 10-dimensional Minkowski spacetime for (type I, type IIA, type IIB, heterotic) superstrings, and supergravity theories. It is also true that 11 is the critical dimension for M theory. Besides these critical dimensions, anomaly cancellation of the ten-dimensional superstring theories requires that the gauge group be $SO(32)$ or $E_8 \times E_8$ [1]. To make contact with a potential four-dimensional Minkowski spacetime effective field theory, one must hence compactify the remaining six-dimensional manifold in a very special way. The simplest approach is to consider the vacuum form for the ten-dimensional spacetime of the form.

$$\mathfrak{M}^{10} = \mathfrak{M}^4 \times X^6 \qquad (1.1)$$

where $\mathfrak{M}^4$ is the usual four-dimensional Minkowski spacetime and $X^6$ is a compactified internal manifold.

Phenomenologically, interesting scenarios arrived when the $E_8 \times E_8$ heterotic string is compactified on a Calabi-Yau 3-fold. This is the outcome of requiring an unbroken $\mathcal{N} = 1$ supersymmetry effective action in the $(3 + 1)$-dimensional Minkowski spacetime [2]. Additionally, the compactification of type IIA or type IIB theories on a Calabi-Yau 3-fold yields $\mathcal{N} = 2$ supersymmetry effective action in the $(3 + 1)$-dimensional Minkowski spacetime. The overall conclusion is that compactifying superstring theory on a Calabi-Yau $n$-fold breaks $\frac{3}{4}$ of the original supersymmetry [4–6]. These results give to Calabi-Yau manifolds a very special position as far as compactification of string theory in its various forms is concerned. The key outcome is that the physics of $\mathfrak{M}^4$ is determined by the geometry of $X^6$ [7].

Calabi-Yau manifolds [complete intersection Calabi-Yau (CICY[1])] are by now understood partially. For instance, $CICY_n$ have a defining configuration matrix, a defining Hodge diamond characterized by Hodge numbers $h^{p,q}$, $(p, q = 1, 2, 3, \ldots)$, the intersection number $d_{rst}(r, s, t = 1, 2, \ldots, h^{1,1})$, and the first Pontrjagin class $p_1(TX^6)$ of the tangent bundle [7]. The Hodge numbers $h^{p,q}$ are the dimensions of the Dolbeault cohomology group of CICYs. As far as $CICY_3$ are concerned, one is left with $h^{1,1}$ and $h^{2,1}$ unspecified. These numbers do not uniquely define the manifold in question, however. There are moduli deformations (changes of shape and size) that do not affect the topology of $CICY_3$. The Hodge number $h^{2,1}$, for instance, represents the number of deformation in the complex structure (shape) of the Calabi-Yau manifold in question. In brief, in any Calabi-Yau manifold, there is basically a $h^{2,1}$-parameter family of topologically equivalent varieties distinct just by their complex structures. On the other hand, $h^{1,1}$ is the number of the Khähler-structure (size) deformation changes of $CICY_3$. These manifolds have the same topology but different shapes or sizes. The pairs $(h^{1,1}, h^{2,1})$ are therefore a natural playground for applying machine learning (ML) techniques as a test bed for regressions [8].

---

[*]Contact author: madyfalaye@gmail.com

[1]We hereafter denote CICY $n$-fold by $CICY_n$.

The application of techniques from ML in analyzing aspects of CICY is nowadays very productive [9–16]; we refer to these articles and references therein for further information. In these approaches, classification and regression, which can be both supervised and unsupervised, are used. In Ref. [17], the enterprise of using ML to address issues of Calabi-Yau manifolds has been put forward. This author gave winning evidence of the utility of machine learning in addressing issues of CICY. The main interests in these computations are using classification or regression techniques to learn the Hodge numbers $h^{p,q}$ when the configuration matrices are taken as the input of the algorithms [18–22]. Several models have been utilized in this prelude and we are learning that most of these models are performing well on the dataset of $CICY_3$ and $CICY_4$. Using the configuration matrix as an input is, in our view, a bit problematic and it might greatly affect the performance of the model building due to the ambiguities of its definition. This can explain somehow the bad performance in learning $h^{3,1}$ of the deep multitask approach in [22]. A verbatim report about the impact of matrix manipulations in the field of machine learning has not yet been written down as of today. Changing two rows or two columns of a matrix does not affect the key properties of that matrix. This operation, however, has far-reaching consequences as far as CICYs are concerned [3].

The main goal of this manuscript is to avoid this plan of attack. In doing so, we consider one of the Hodge numbers as an input and machine learn the other one. In other words, we learn the Hodge number $h^{2,1}$ in term of $h^{1,1}$. This is a very interesting problem; the two numbers represent the number of vector multiplets $(h^{1,1})$ and the number of hypermultiplets $(h^{2,1} + 1)$ as far as one is compactifying IIA string theory on a Calabi-Yau manifold [4].

The shape and the size of a manifold are naturally related; the primary objective of this article is to access the nature of this connection for $CICY_3$. This is the domain of the regression techniques in machine learning and there are several algorithms to this end. In this work, we will apply Gaussian process regression (Gausspr) [23], kernel support vector regression (KSVM) [24–26], the random forest (RF) [27,28], extreme gradient boosting (Xgboost) [29], and the generalized linear model (Glm) [30]. A comparative study of the performance of these regressions is discussed in this paper. We also conduct a clustering of CICY into three groups. This is an alternative way of studying CICY by ML. Our results supply supporting facts about the usefulness of ML in dealing with $CICY_3$. The remaining parts of this paper are divided as follows: In Sec. II, facts on CICY and aspects of string compactification on a CICY are presented. In Sec. III, we apply regression algorithms to learning $h^{2,1}$ in terms of $h^{1,1}$ and we describe the hyperparameters of the chosen regressions. The results of these regressions that apply to $CICY_3$ are laid out and discussed in Sec. IV. Finally, Sec. V is concentrated on conclusions and the possible extension of this work.

## II. COMPLETE INTERSECTION CALABI-YAU MANIFOLDS AND STRING COMPACTIFICATION

Understanding the structure of the internal manifold in superstring compactification is crucial in string theory; special attention has been paid to the study of CICY. Hence, we begin with the presentation of some useful concepts about $CICY_n$, which an overview of string compactification on these manifolds will follow. It is interesting to note that this Calabi-Yau compactification is phenomenologically very instructive.

### A. Facts about the complete intersection of Calabi-Yau manifolds: CICY

In this subsection, we review the basic concepts of CICY. The standard references which we will highlight are Refs. [3,31,32,34,35], where the basic properties of CICYs have been derived. First of all, let the ambiant space $\mathbb{M} = \mathbb{P}_1^{n_1} \times \mathbb{P}_2^{n_2} \times \ldots \times \mathbb{P}_m^{n_m}$ be a product of $m$ complex projective spaces of complex dimensions $n_i$, respectively.[2] In addition, let $\mathcal{P}^\alpha(z_r^\mu), (\alpha = 1, 2, \ldots, K)$ be $K$ homogeneous holomorphic polynomials such that if $z_r^\mu$ is the homogeneous coordinates of $\mathbb{P}_r^{n_r}$, then $\mathcal{P}^\alpha(\lambda z_r^\mu) = \lambda^{q_\alpha^r} \mathcal{P}^\alpha(z_r^\mu)$. The sets $q_\alpha^r$ are called the homogeneous degrees of the polynomial $\mathcal{P}^\alpha(z_r^\mu)$ with respect to the coordinates of $\mathbb{P}_r^{n_r}$. Finally, let $\mathcal{X}^\alpha \colon \mathcal{P}^\alpha(z_r^\mu) = 0$ be the zero locus of $\mathcal{P}^\alpha(z_r^\mu)$, and then we have the following fact.

*Fact 1*: The manifold $\mathbb{X} = \mathcal{X}^1 \cap \mathcal{X}^1 \cap \mathcal{X}^2 \cap \ldots \cap \mathcal{X}^K \subset \mathbb{M}$ (which is a complete intersection of the hypersurfaces $\mathcal{X}^\alpha$) is a Calabi-Yau manifold[3] if and only if we have

$$\sum_{\alpha=1}^{K} q_\alpha^r = n_r + 1, \quad r = 1, \ldots, m. \quad (2.1)$$

These concepts about a CICY can be encoded into the so-called *configuration matrix* denoted by

$$\mathbb{X} = \begin{bmatrix} \mathbb{P}^{n_1} & q_1^1 & \cdots & q_K^1 \\ \mathbb{P}^{n_2} & q_1^2 & \cdots & q_K^2 \\ \vdots & \vdots & \ddots & \vdots \\ \mathbb{P}^{n_m} & q_1^m & \cdots & q_K^m \end{bmatrix} = \begin{bmatrix} n_1 & q_1^1 & \cdots & q_K^1 \\ n_2 & q_1^2 & \cdots & q_K^2 \\ \vdots & \vdots & \ddots & \vdots \\ n_m & q_1^m & \cdots & q_K^m \end{bmatrix}, \quad q_a^r \in \mathbb{Z}_{\geq 0}. \tag{2.2}$$

---
[2]We abbreviate $\mathbb{CP}$ by $\mathbb{P}$.
[3]Vanishing of its first Chern Class.

The dimension of the CICY manifold $\mathbb{X}$ is derived from Eq. (2.3).

$$\sum_{r=1}^{m} n_r = dim_{\mathbb{C}}\mathbb{X} + K. \tag{2.3}$$

Equation (2.1) along with (2.3) signal the finiteness of $CICY_n$ for any $n$. More importantly, the Hodge numbers can be derived from the configuration matrix [3,31–34].

*Fact 2*: The number of $CICY_n$ is finite due to Eqs. (2.1) and (2.3). For instance, by considering one single complex projective space $\mathbb{P}^n$ and $K$ polynomials intersecting on it, we have

$$n = dim_{\mathbb{C}}\mathbb{X} + K \quad \text{and} \quad \sum_{\alpha=1}^{K} q_\alpha^r = n+1 \quad \text{implying that} \quad \sum_{\alpha=1}^{K} q_\alpha^r = dim_{\mathbb{C}}\mathbb{X} + K + 1. \tag{2.4}$$

As an illustration, for $dim_{\mathbb{C}}\mathbb{X} = 4$ and $K = 1$, one gets $(n, q) = (5, 6)$; this manifold is denoted by $[5||6]$. In the same case for $K = 2$, one may get $[6||34]$ and $[6||25]$. In these examples, we are requiring that $q_\alpha^r \geq 2$. The Fermat quintic equation in $\mathbb{P}^4$ denoted by $[4||5]$ as well as the solution space for the intersection of a quartic and a quadratic in $\mathbb{P}^5$, $[5||42]$, are interesting examples of $CICY_3$. More importantly, the quotient of the Tian-Yau manifold [36,37] (2.5) by a freely acting $\mathbb{Z}_3$ is of capital interest in $CICY_3$ compactification. It corresponds to three generations of particles in the compactified string theory, which is an interesting result [38–40]. Note that the Tian-Yau manifold itself is an example of noncompact CICY 3-folds. Nevertheless, the quotient of the Tian-Yau manifold by a freely acting $\mathbb{Z}_3$ is a compact 3-fold.

$$\mathbb{X}_{TY} = \begin{bmatrix} 3 & 1 & 3 & 0 \\ 3 & 1 & 0 & 3 \end{bmatrix}. \tag{2.5}$$

Most of the interesting manifolds are obtained by taking the quotient of a manifold of $SU(3)$ holonomy by one of its discrete freely acting symmetry groups. In doing so, one obtains a phenomenologically relevant Euler number [38].

*Fact 3*: For any $CICY_n$ the Hodge number $h^{p,q}$ satisfies the complex conjugation and the Hodge star or Poincaré duality:

$$h^{p,q} = h^{q,p} \quad \text{and} \quad h^{n-p,n-q} = h^{p,q}. \tag{2.6}$$

Using these properties, one easily finds that the only unspecified Hodge numbers for $CICY_3$ are $h^{1,1}$ and $h^{2,1} = h^{1,2}$. This is because the defining holomorphic top form for $CICY_3$ is $h^{3,0} = 1$, which sets the Hodge diamond corners. On the other hand, the simple connectedness of $CICY_3$ implies $h^{1,0} = 1$, setting the Hodge diamond edges. Consequently, the Euler characteristic $\chi$ of any $CICY_3$ is then given by the formula

$$\chi = 2(h^{1,1} - h^{2,1}). \tag{2.7}$$

### B. String compactification on a Calabi-Yau manifold

Having presented a few facts about Calabi-Yau complete intersections, we are now in a position to proceed by looking at string compactification on these manifolds. The standard references on this aspect are Refs. [2,4–6,41,42]. We will start by reviewing aspects of string compactification that are relevant for $CICY_3$ manifolds. This restriction does not alter the conclusion of this manuscript as far as the dataset of $CICY_3$ is concerned. When various string theories are compactified on a $CICY_3$, one finds an interesting four-dimensional theory with $\mathcal{N} = 2$ supersymmetry.

The four-dimensional theory with $\mathcal{N} = 2$ supersymmetry, from the compactification of superstring theory on a Calabi-Yau 3-fold, has interesting aspects. For instance, the $\mathcal{N} = 2$ supermultiplets are divided into Abelian vector multiplets and hypermultiplets. For IIA and IIB string compactifications on a $CICY_3$, the numbers of these multiplets are given in Table I.

TABLE I. The division of $\mathcal{N} = 2$ supermultiplets.

| String Compactified on a CICY$_3$ | $\mathcal{N} = 2$ Supermultiplets | |
| --- | --- | --- |
| Type | Vector multiplets | Hypermultiplets |
| IIA | $h^{1,1}$ | $h^{2,1} + 1$ |
| IIB | $h^{2,1}$ | $h^{1,1} + 1$ |

If two CICY$_3$ $\mathbb{X}^1$ and $\mathbb{X}^2$ are such that $h^{1,1}(\mathbb{X}^1) = h^{2,1}(\mathbb{X}^2)$, then the manifolds $\mathbb{X}^1$ and $\mathbb{X}^2$ are said to be mirror. One immediate consequence of (Table I) is that type IIA superstring theory compactified on $\mathbb{X}^1$ is equivalent to type IIB superstring theory compactified on $\mathbb{X}^2$. The mirror symmetry also implies the self-duality nature of the space of string theory vacua. The implication of mirror symmetry in what follows is an implicit connection between learning $h^{2,1}$ in terms of $h^{1,1}$ and learning $h^{1,1}$ in terms of $h^{2,1}$. One natural intent of this manuscript is to clarify this issue.

Upon compactification of the $E_8 \times E_8$ heterotic string on a CICY$_3$ a few things happen. The most noticeable is that one of the $E_8$ is broken to $E_6$ with the following decomposition of its 248-dimensional representation:

$$E_8 \longrightarrow SU(3) \times E_6$$
$$248 \longrightarrow (1, 78) \oplus (3, 27) \oplus (\bar{3}, \overline{27}) \oplus (8, 1). \quad (2.8)$$

From the decomposition (2.8), we are thus able to put all the massless fields into the 27 representation of $E_6$ by using a grand unification theory based on $E_6$. The generation one gets is half the Euler characteristic of the Calabi-Yau manifold [2,31]. This outcome is generally true for any Calabi-Yau compactification of the heterotic string theory.

## III. REGRESSION ALGORITHMS FOR LEARNING $h^{2,1}$ IN TERMS OF $h^{1,1}$

The basic purpose of regression algorithms is to find the best hypothesis function $\mathfrak{f}$ which relates some input values to the output ones [43]. When the input is one variable as in our case, then the regressions are called univariate or simple. This is in contrast to multivariate regression scenarios. The regression can be either linear or nonlinear depending on the nature of the function $\mathfrak{f}$ used. In this work, we model our regressions by the formula given by

$$\hat{h}^{2,1} = \mathfrak{f}(h^{1,1}) \quad (3.1)$$

where the output of the regression, the predicted value, is $\hat{h}^{2,1}$ and the target output is $h^{2,1}$. The error of the predictions is the difference between the true target value $h^{2,1}$ and predicted target value $\hat{h}^{2,1}$. The relationship (3.1) is known as a simple regression in contrast to multiple regression or multivariate regression. The true value $h^{2,1}$ is related to the model prediction $\hat{h}^{2,1}$ by the relation

$$h^{2,1} = \hat{h}^{2,1} + \epsilon, \quad \text{where } \epsilon \text{ is the error.} \quad (3.2)$$

Different approaches try to minimize the value of $\epsilon$ by using various minimization techniques. Before the application of the various regressions' techniques, one has to divide the dataset into a train set (80%) and a validation set (20%). Sometimes, the data are separated into a train set, a validation set, and a test set, but we will not follow this option in our work. The performance of a chosen model is measured by a few statistical parameters such as the mean square error (MSE), the root mean square error (RMSE), the mean absolute error (MAE), the Pearson correlation coefficient ($R^2$), and so on. One quantity that signals the underestimation (negative value of the bias) or overestimation (positive value of the bias) for the output by a given model is the bias. For these parameters, the formula is given in Refs. [19,43,44]

$$\text{MSE} = \frac{1}{N} \sum_{i=1}^{N} (h_i^{2,1} - \hat{h}_i^{2,1})^2$$
$$\text{RMSE} = \sqrt{MSE},$$
$$\text{MAE} = \frac{1}{N} \sum_{i=1}^{N} |h_i^{2,1} - \hat{h}_i^{2,1}|$$
$$R^2 = 1 - \frac{\sum_{i=1}^{N} (h_i^{2,1} - \hat{h}_i^{2,1})^2}{\sum_{i=1}^{N} (h_i^{2,1} - \bar{h}^{2,1})^2}$$
$$\text{BIAS} = \frac{1}{N} \sum_{i=1}^{N} (\hat{h}_i^{2,1} - h_i^{2,1}) \quad (3.3)$$

where in these equations, $N$ denotes the number of elements in the set (train set or validation set) under consideration and $\bar{h}^{2,1}$ represent the mean of $h^{2,1}$.

The best model must have the highest value for $R^2$ and the smallest value for the other parameters following the measure definitions in Eq. (3.3). Perfect learning gives $R^2 = 1$ and the remaining parameters in Eq. (3.3) vanish. To compute these parameters, we use the aforementioned algorithms which we will now outline in the remaining part of this section. The regression techniques use some parameters that are often called hyperparameters; the suitable values of these hyperparameters are also given.

### A. The hyperparameters of the regressions techniques

The first regression technique we applied is the Gausspr. This is currently a famous regression technique used in the field of machine learning. The full theory goes beyond the scope of this manuscript. The necessary information about Gausspr can be found in Refs. [45,46]. In Gausspr the hypothesis function is a Gaussian function and it is characterized by some hyperparameters. To implement the algorithm of Gausspr, we used the software R and the package kernlab [47]. The hyperparameters of Gausspr are (var = 0.15, kpar = automatic, cross−validation = 5, and kernel = polydot), the list of hyperparameters (kernel parameters) are abbreviated in kpar, and the initial noise variance is denoted by var. This regression is selected due to its nonparametric Bayesian approach by using the so-called Gaussian process [48].

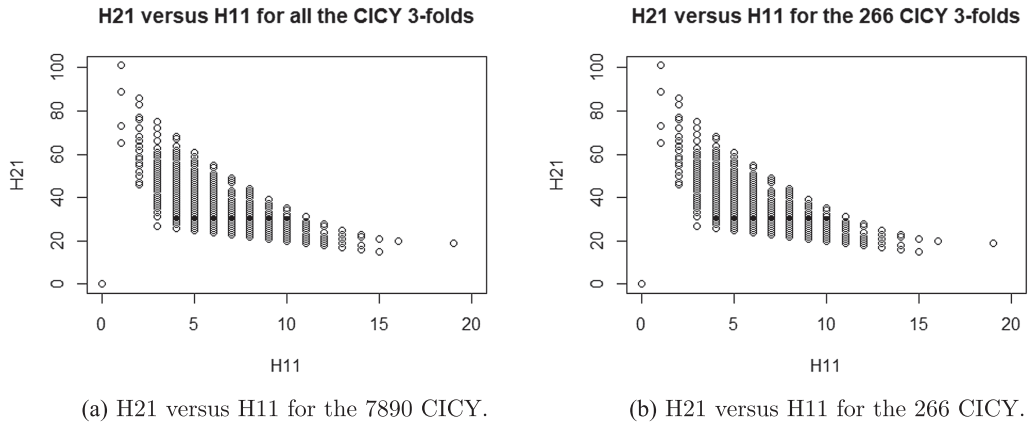

FIG. 1. The plot of H21 versus H11.

The second algorithm is the so-called KSVM. In this approach, one uses a kernel (function) to transform the data points into a feature space. Then, an $\epsilon$-insensitive error function is defined to form a hyperplane within the feature space. The data points which lie on the hyperplanes are called the support vectors. Again, the full mathematics are omitted and can be consulted in Ref. [45]. To implement KSVM, we also used R and the kernlab package. The hyperparameters used in KSVM are (kernel = polydot, kpar = automatic, C = 4, nu = 0.1, epsilon = 0.1, cross−validation = 0, cache = 40, and the tolerance of termination criterion tol = 0.15). Here nu sets the upper bound on the training error and the lower bound on the fraction of data points to become support vectors; C is the cost of constraints violation.

The next regression is RF [27,28] which uses the concepts of decision trees. For its implementation, we used the package randomforest [49]. Its basic parameters are the number of trees to grow (ntree = 25), the number of variables randomly sampled as candidates at each split (mtry = 2), and the minimum size of terminal nodes (nodesize = 3).

The fourth regression applied is Xgboost. This is a perfect regression algorithm. The package Xgboost [50] has been used for its execution. It has two hyperparameters: the max number of boosting iterations (nrounds = 69) and the maximum depth of a tree (max depth = 3). The last regression does not have hyperparameters to set in. We will survey the dataset of $CICY_3$ before the application of these techniques of regressions.

The results of these regressions are compared to the Glm [51,52] and to the results of the traditional partial least square (PLS) [53].

### B. Datasets of $CICY_3$ and their splitting into train and validation sets

The complete datasets of $CICY_3$ are composed of 7890 different topological types of $CICY_3$. However, there are only 266 distinct pairs of $(h^{1,1}, h^{2,1})$. The plot of the Hodge number $h^{2,1}$ (H21) in the function of $h^{1,1}$ (H11) is given in Fig. 1. Several values of $h^{2,1}$ are mapped to a single value of $h^{1,1}$. Consequently, the machine learning of $h^{2,1}$ in terms of $h^{1,1}$ is more challenging. This is virtually the same trouble one encounters in learning $h^{2,1}$ or $h^{1,1}$ by using the *configuration matrices* as input of the algorithms. We divide the 7890 $CICY_3$ into 6312 (80%) varieties used for training the models and 1578 (20%) manifolds for validating the algorithms.

### IV. RESULTS AND DISCUSSIONS

The purpose of this section is to show the different results and discuss them in some detail. The first result is Table II which contains the values of the statistical parameters in Eq. (3.3) except the values of MAE.

From this table, we observe that the Gaussian process regression is the most powerful in terms of performance (lowest value of RMSE). It is then followed by the kernel support vector machine, the random forest, the extreme gradient boost, and the generalized linear model. We surprisingly discovered that for the data set, the traditional PLS has more or less the same statistical parameters as the generalized linear model. One noticeable thing about Table II is that all of the applied techniques overestimate the Hodge number $h^{2,1}$ for the validation which is visible from the bias value.

Next, we have the graphs of the train lines of Gausspr in Fig. 2 and KSVM in Fig. 3. These plots show that the two algorithms precisely predicted the Hodge number $h^{2,1}$.

TABLE II. Few statistical Parameters of the applied regressions techniques.

| Regression | Validation | | | Train | | |
|---|---|---|---|---|---|---|
| | $R^2$ | RMSE | BIAS | $R^2$ | RMSE | BIAS |
| Gausspr | 0.9999999995 | 0.0002895011 | 3.754119E-6 | 0.9999999994 | 0.0002854348 | $-6.001535E-10$ |
| KSVM | 0.9999982 | 0.3527558 | 0.3439119 | 0.9999980 | 0.3515819 | 0.3432654 |
| RF | 0.9984065 | 0.35785376 | 0.009894381 | 0.999977440 | 0.040472929 | $-0.000273553$ |
| Xgboost | 0.7048157 | 4.7934372 | $-0.1547063$ | 0.6921789 | 4.72444 | 2.958037E-7 |
| Glm | 0.5192372 | 6.123511 | $-0.07941958$ | 0.4972947 | 6.037509 | $-1.163628e-13$ |
| PLS | 0.5192372 | 6.1235110 | $-0.07941958$ | 0.4972947 | 6.0375093 | $-7.490853E-16$ |

In all the graphs below, $h^{2,1}$ is denoted by $H21$ and $H11$ stands for $h^{1,1}$. It is interesting to note that these regression techniques can be utilized to a great extent in the field of machine learning Calabi-Yau manifolds.

The graphs for the regression lines of random forest and extreme gradient boost are depicted in Figs. 4 and 5, respectively.

We notice that the random forest performs badly when the Hodge number $h^{2,1}$ is bigger. As for the predictions from the extreme gradient boost, one observes that they are dispersed almost symmetrically around the regression line. The roots of this poor performance of the extreme gradient boost are unclear. We therefore advise avoiding this regression algorithm in the field under consideration. It is a very good minimization approach, but not in the flatland of machine learning of $CICY_3$. In stark contrast, however, we do have the superposition of the original and the predicted values of $h^{2,1}$ in Fig. 6(a), which shows that the prediction is not as bad as it might first appear in Fig. 5(a).

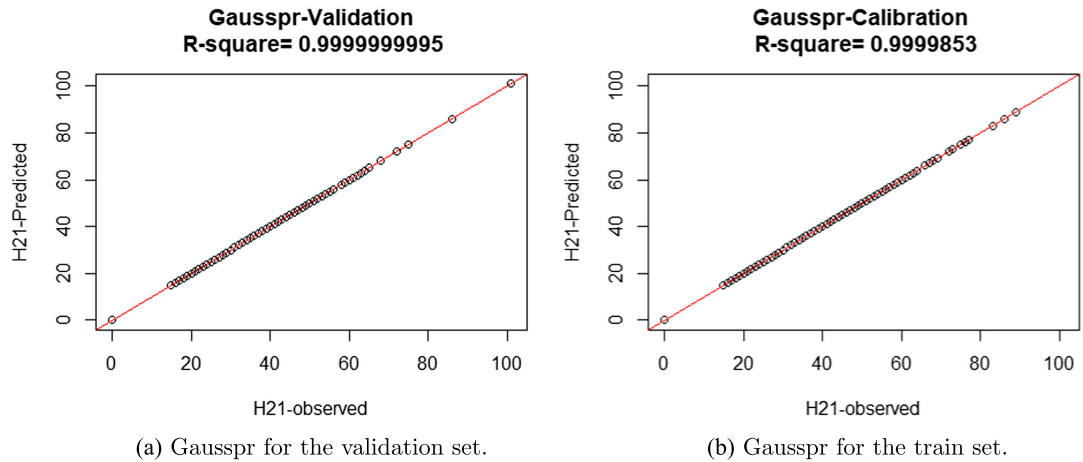

(a) Gausspr for the validation set.           (b) Gausspr for the train set.

FIG. 2.   The predictions from Gausspr.

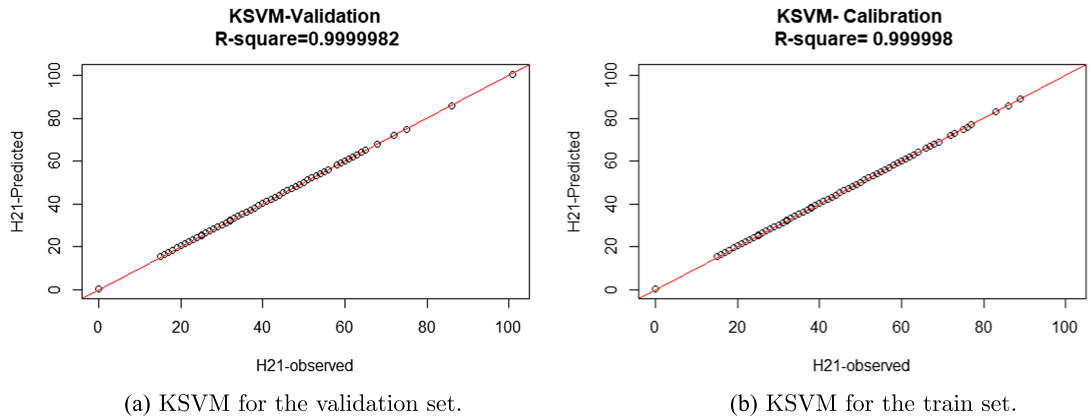

(a) KSVM for the validation set.           (b) KSVM for the train set.

FIG. 3.   The predictions from kernel support vector machine.

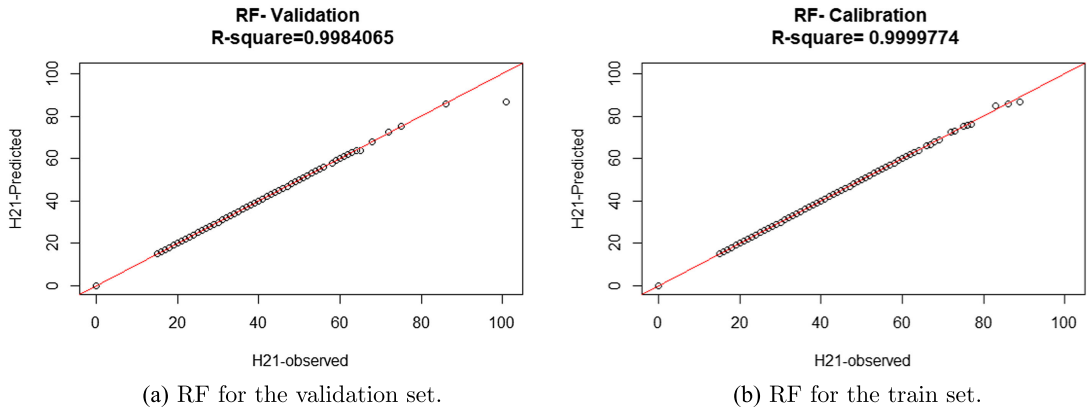

(a) RF for the validation set.  (b) RF for the train set.

FIG. 4. The predictions from the random forest.

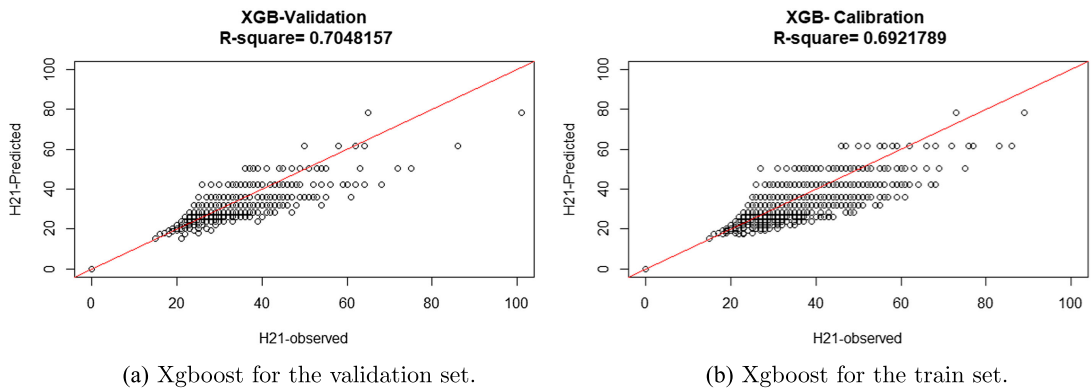

(a) Xgboost for the validation set.  (b) Xgboost for the train set.

FIG. 5. The predictions from the extreme gradient boost.

The last plots of the regression lines are the ones for the generalized linear model and the partial least square, which are given in Figs. 7 and 8, respectively.

These plots reveal that the glm and the PLS models are definitely not suitable for addressing issues regarding the

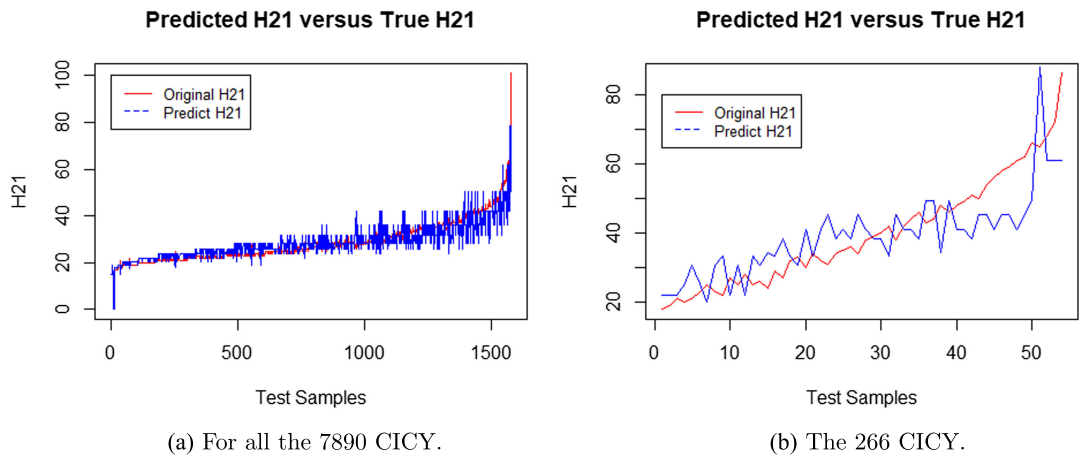

(a) For all the 7890 CICY.  (b) The 266 CICY.

FIG. 6. The predictions superposed to its original values of $h^{2,1}$ from Xgboost.

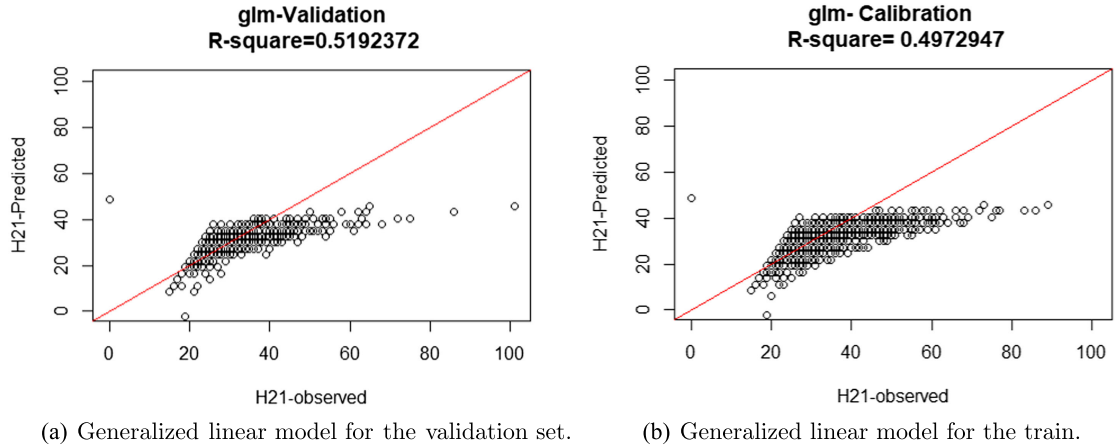

(a) Generalized linear model for the validation set.  (b) Generalized linear model for the train.

FIG. 7. The predictions from the generalized linear model.

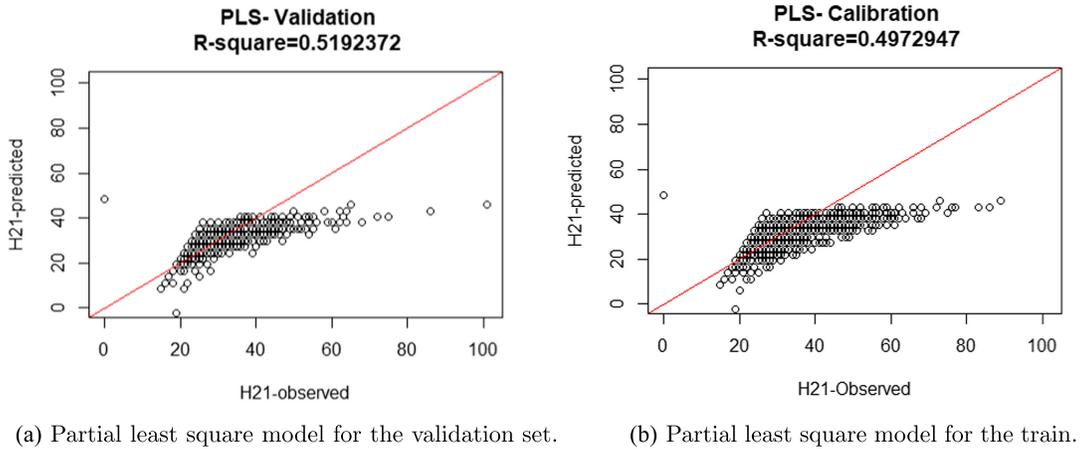

(a) Partial least square model for the validation set.  (b) Partial least square model for the train.

FIG. 8. The predictions from the partial least square model.

application of machine learning of Calabi-Yau manifolds. Equally predictable is that the Xgboost, glm, and PLS are underestimating the values of the Hodge number $h^{2,1}$ which can be seen from the plots of the graphs of these models as well as in the values of their bias in Table II above.

In the field of machine learning, an important concept is the partitioning clustering of the data points into clusters [54,55]. In this article, we apply three types of clustering: the K-means clustering, the partitioning around medoids (PAM) clustering and the clustering large applications (clara) method. The former is sensitive to outliers whereas PAM is less sensitive to outliers. The figures of these clustering are shown in Figs. 9 and 10 below. In the jargon of data scientists, PAM is also known as K-medoids clustering [56,57]. These popular algorithms for clustering are presented for future tasks on unsupervised machine learning approaches to $CICY_3$. Analyzing the usefulness of clustering $CICY_3$ into several groups and the physics associated to each group needs further probes and we postpone these investigations for future considerations. Notwithstanding, we detect a striking similarity between Figs. 9(b) and 9(c). The number of clusters [Fig. 9(a)] is selected according to the proposed method of Ref. [58].

The interpretation of Fig. 10(b) is as follows: they are 3334 data points in the first cluster, 3007 data points in the second cluster, and, finally, 1549 data points in the last cluster. In addition, the values of $s_i$ indicate that a reasonable structure has been found in clusters 1 and 2, whereas, the structure found in cluster 3 could be artificial [59].

### A. Regressions techniques and the number of generations

In the context of Calabi-Yau compactification, the number of particle generations in the four-dimensional world is given by relation [2].

$$\text{Number of Generation} = \frac{1}{2}|h^{1,1} - h^{2,1}|. \quad (4.1)$$

Examples of Calabi-Yau manifolds having linear relations like Eq. (4.1) are scarce (one can look at for instance,

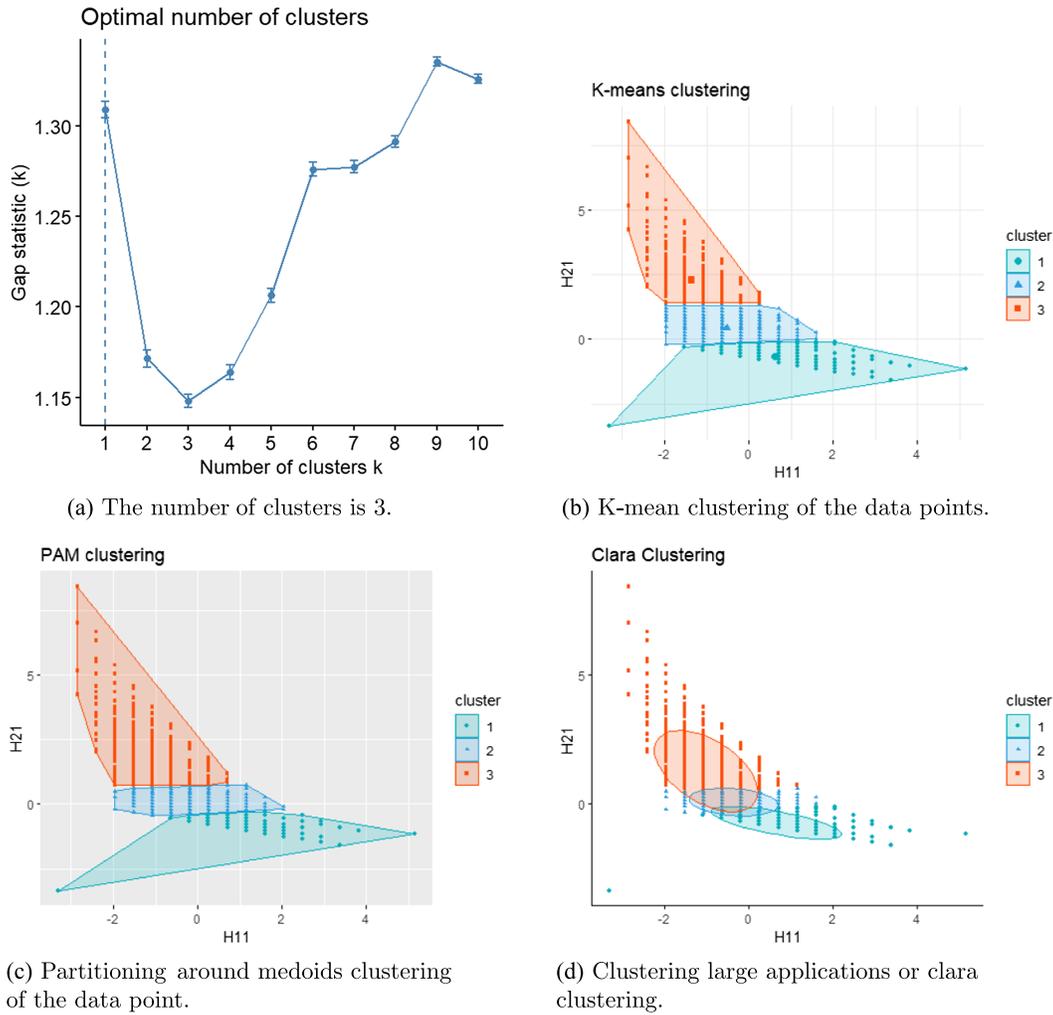

FIG. 9. The three types of clustering of the data points.

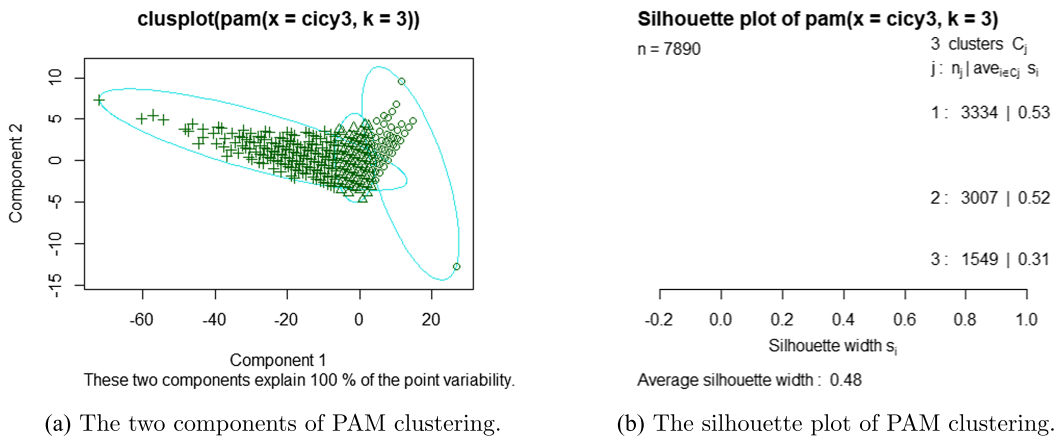

FIG. 10. Silhouette and components plots of PAM clustering.

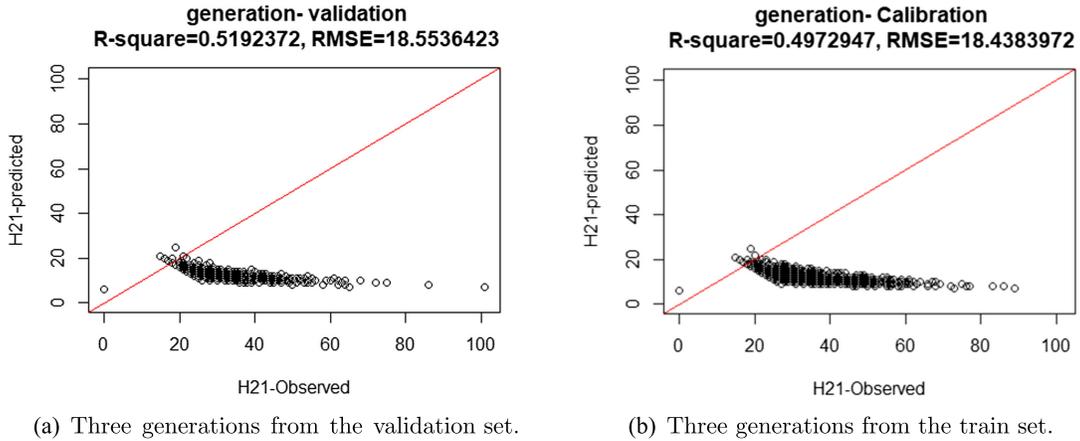

FIG. 11. The predictions from $\mathfrak{f}(h^{1,1}) = h^{1,1} + 6$.

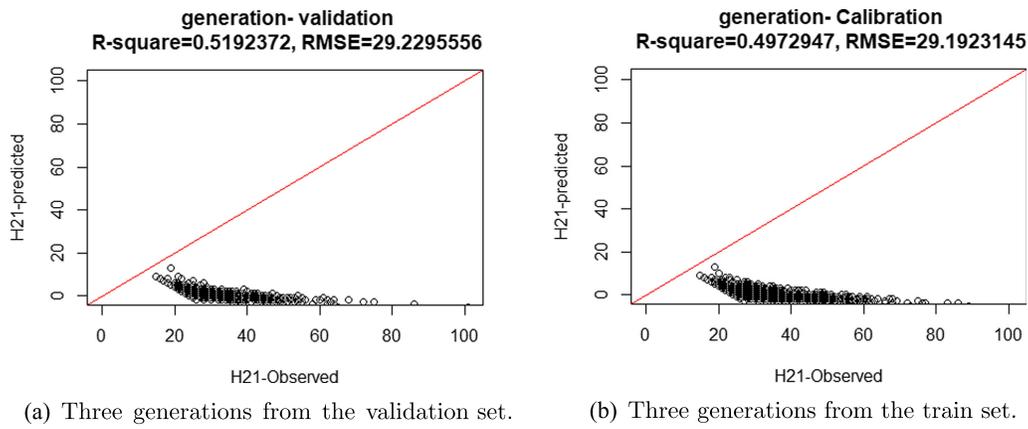

FIG. 12. The predictions from $\mathfrak{f}(h^{1,1}) = h^{1,1} - 6$.

Tables 7 and 9 of [60]). In order to have three generations in our regression models, we use Eq. (4.1) and the number of particle generations becomes

$$\text{Number of Generation} = \frac{1}{2}|h^{1,1} - \hat{h}^{2,1}| = 3,$$

implying that $\mathfrak{f}(h^{1,1}) = h^{1,1} \pm 6$. (4.2)

The number of manifolds satisfying Eq. (4.2) is shown in Figs. 11 and 12. The manifolds on the red line are the ones that admit three-generations criterion.

We observe that $h^{2,1} \approx \mathfrak{f}(h^{1,1}) = h^{1,1} - 6$ does not match any CICY 3-folds and $h^{2,1} \approx \mathfrak{f}(h^{1,1}) = h^{1,1} + 6$ matches very few.

A similar analysis can be carried out in replacing the 7890 complete intersection Calabi-Yau manifolds by the datasets of the 266 distinct pairs of $(h^{1,1}, h^{2,1})$ Calabi-Yau

TABLE III. Statistical parameters for models applied to the 266 $CICY_3$.

| | Validation | | | Train | | |
|---|---|---|---|---|---|---|
| Regression | $R^2$ | RMSE | MAE | $R^2$ | RMSE | MAE |
| Gausspr | $1.000000e + 00$ | 8.913495E-5 | 7.209163E-5 | $1.000000e + 00$ | 9.713904E-5 | 7.130115E-5 |
| KSVM | 0.9998109 | 0.6911974 | 0.5557933 | 0.9996605 | 0.7001122 | 0.5539415 |
| RF | 0.9883554 | 1.7488026 | 1.0432899 | 0.9888899 | 1.7384071 | 0.8420099 |
| Xgboost | 0.6094061 | 9.7423305 | 8.0058412 | 0.6355857 | 9.0960680 | 7.1641955 |
| Glm | 0.6084452 | 10.1644505 | 8.2209245 | 0.459558 | 11.077214 | 8.130806 |
| PLS | 0.6084452 | 10.1644505 | 8.2209245 | 0.459558 | 11.077214 | 8.130806 |

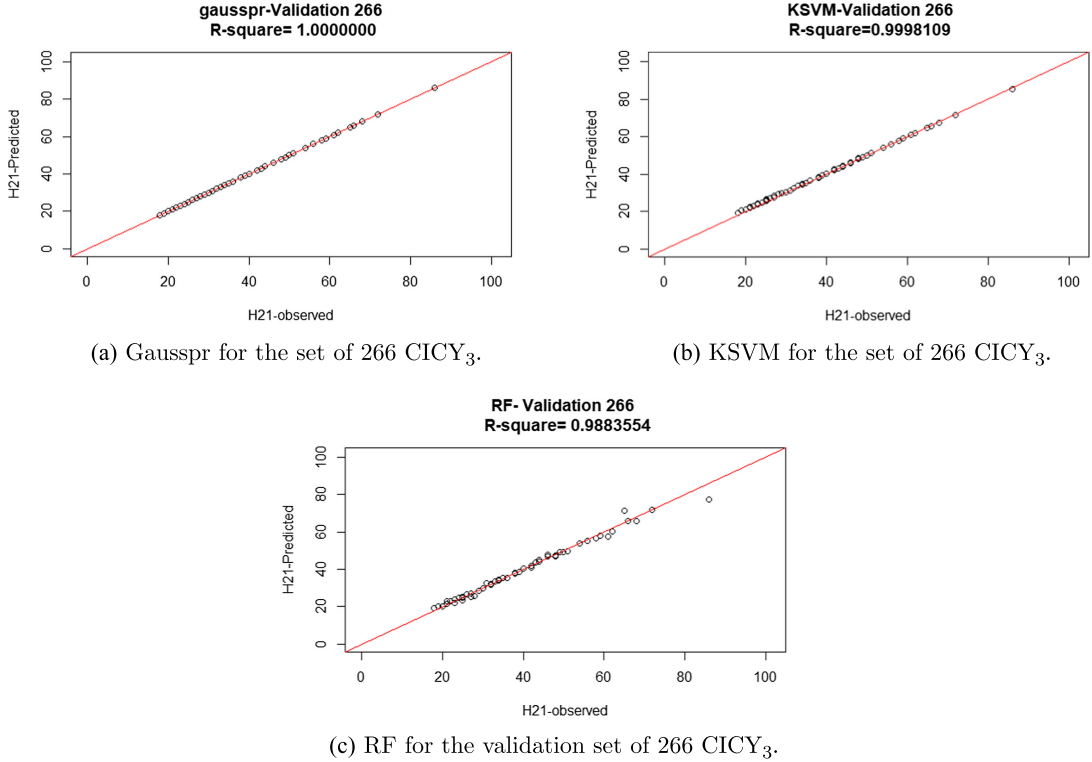

(a) Gausspr for the set of 266 CICY$_3$.

(b) KSVM for the set of 266 CICY$_3$.

(c) RF for the validation set of 266 CICY$_3$.

FIG. 13. The predictions for the validation set of 266 CICY$_3$.

manifolds. The (80%, 20%) division of the dataset into the training sets and the validation sets results in 212 samples for calibrating the models and 54 samples for validating them. The statistical parameters for the above regressions are indicated in Table III.

Not surprisingly, the order of performance for the regression techniques remains the same. The most powerful one is always Gausspr which is followed by KSVM. We then have RF at the third position in terms of performance. The plots of the regression lines of the validation sets for these three algorithms are depicted in Fig. 13.

A glance at these figures reveals the excellent performance of these three regression techniques. It is apparent from Fig. 13(c) that random forest has to be taken with caution when the values of $h^{2,1}$ is higher than 50. These results resume the learning of $h^{2,1}$ as the output of the regressions when $h^{1,1}$ is taken as the input.

Likewise, one can instead take $h^{2,1}$ as the input of the algorithms and learn $h^{1,1}$ which is an interesting approach. The results for this option are resumed in Table IV and in Fig. 14.

We see that this alternative way of machine learning Calabi-Yau manifolds is also possible in practice. More importantly, the performance order of the regression techniques remains unchanged.

TABLE IV. Few statistical Parameters for learning $h^{1,1}$ in terms of $h^{2,1}$.

| | Validation | | | Train | | |
|---|---|---|---|---|---|---|
| Regression | $R^2$ | RMSE | BIAS | $R^2$ | RMSE | BIAS |
| Gausspr | 1.000000e + 00 | 7.395731E-5 | 6.503978E-7 | 1.000000e + 00 | 7.540186E-5 | 9.016338E-11 |
| KSVM | 0.9999217 | 0.03692016 | 0.01780263 | 0.9999231 | 0.03759887 | 0.01803469 |
| RF | 0.9999994 | 0.001744083 | 7.604563E-5 | 0.999998 | 0.00314419 | 5.703422e−05 |
| Xgboost | 0.7529686 | 1.121953 | −0.03815101 | 0.7359613 | 1.155875 | −2.051232E−6 |
| Glm | 0.5192372 | 1.564346 | −0.01375948 | 0.4972947 | 1.5949 | −3.607428e−14 |

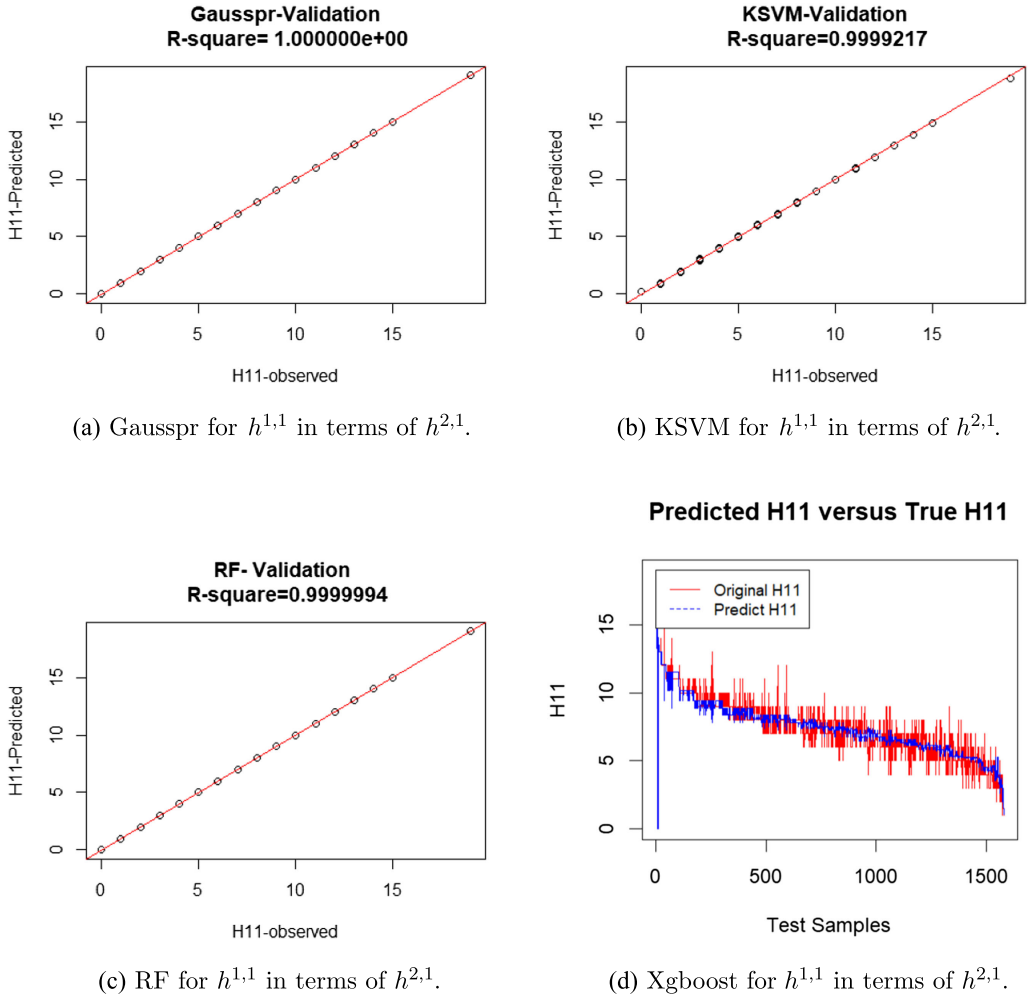

FIG. 14. The predictions for $h^{1,1}$ in terms of $h^{2,1}$.

## V. CONCLUSIONS

In this paper, by using several regression techniques, we analyzed the dataset of 7890 complete intersection Calabi-Yau manifolds of complex dimension three. First of all, we took the Hodge number $h^{1,1}$ as the input of the algorithms and machine learned the Hodge number $h^{2,1}$. The performance of these univariate regressions is measured by some statistical parameters which are reported in Tables II and III. We have also machine learned the Hodge number $h^{1,1}$ in terms of $h^{2,1}$ and the outcome is reported in Table IV. By looking at the values of RMSE, one can see that the Gaussian process regression is the most suitable regression technique for the datasets of complete intersection Calabi-Yau 3-folds. It is interesting to note that the extreme gradient boost may encounter some issues when applied to $CICY_3$. The clustering into three groups of the complete intersection of Calabi-Yau 3-folds is also carried out in this paper. This approach can be used to machine learn the 32 families of $CICY_3$ having nontrivial torsion in the integral cohomology group [61]. Similar analysis can be applied to the data set of complete intersection Calabi-Yau four-folds [22] and five-folds [14]. Our work can also be extended to analyze topological equivalent manifolds which is a challenging problem in itself [64].

By taking $h^{2,1}$ as the input and learning $h^{1,1}$ as the output of the regressions algorithms, we observed that the classification of the regression techniques in terms of performance is not affected. This is not quiet surprising, since type IIA and type IIB string theories are related by dualities [62,63].

## ACKNOWLEDGMENTS

I am grateful to H. Sangare for useful and stimulating discussions about the different techniques employed in this manuscript. Especially, extreme gradient boost was suggested to me by him. It is a great pleasure for me to acknowledge fruitful conversations with G. Thompson.

## DATA AVAILABILITY

The dataset used in this work is available and we extracted the Hodge numbers from the text file [65].